\documentclass[article,superscriptaddress,aps,twocolumn,pra,longbibliography]{revtex4-1}
\usepackage{dcolumn}
\usepackage{bm}
\usepackage{epsfig}
\usepackage{graphicx}
\usepackage{amssymb,amsmath,amsbsy,amsgen,amsfonts,amsthm,mathtools}    
\usepackage{mathrsfs}
\usepackage{latexsym}
\usepackage{array}
\usepackage{color}
\usepackage{amstext}
\allowdisplaybreaks[1]
\usepackage{txfonts}
\usepackage{pifont}
\usepackage{verbatim}
\usepackage{hyperref}

\newcommand{\bra}[1]{\left\langle{#1}\right\vert}
\newcommand{\ket}[1]{\left\vert{#1}\right\rangle}

\DeclareSymbolFont{symbols}{OMS}{cmsy}{m}{n}

\begin{document}
\title{Optimal circular dichroism sensing with quantum light: Multi-parameter estimation approach}

\author{Christina Ioannou}
\affiliation{Institute of Theoretical Solid State Physics, Karlsruhe Institute of Technology, 76131 Karlsruhe, Germany}

\author{Ranjith Nair}
\affiliation{School of Physical and Mathematical Sciences, Nanyang Technological University, 637371, Singapore}
\affiliation{Complexity Institute, Nanyang Technological University, 637335, Singapore}

\author{Ivan~Fernandez‐Corbaton}
\affiliation{Institute of Nanotechnology, Karlsruhe Institute of Technology, 76021 Karlsruhe, Germany}

\author{Mile Gu}
\affiliation{School of Physical and Mathematical Sciences, Nanyang Technological University, 637371, Singapore}
\affiliation{Complexity Institute, Nanyang Technological University, 637335, Singapore}
\affiliation{Centre for Quantum Technologies, National University of Singapore, 117543, Singapore}

\author{Carsten Rockstuhl}
\affiliation{Institute of Theoretical Solid State Physics, Karlsruhe Institute of Technology, 76131 Karlsruhe, Germany}
\affiliation{Institute of Nanotechnology, Karlsruhe Institute of Technology, 76021 Karlsruhe, Germany}

\author{Changhyoup Lee}
\email{changhyoup.lee@gmail.com}
\affiliation{Institute of Theoretical Solid State Physics, Karlsruhe Institute of Technology, 76131 Karlsruhe, Germany}
\date{\today}


\begin{abstract}
The measurement of circular dichroism (CD) has widely been exploited to distinguish the different enantiomers of chiral structures. It has been applied to natural materials (e.g. molecules) as well as to artificial materials (e.g. nanophotonic structures). However, especially for chiral molecules the signal level is very low and increasing the signal-to-noise ratio is of paramount importance to either shorten the necessary measurement time or to lower the minimum detectable molecule concentration. 
As one solution to this problem, we propose here to use quantum states of light in CD sensing to reduce the noise below the shot noise limit that is encountered when using coherent states of light. 
Through a multi-parameter estimation approach, we identify the ultimate quantum limit to precision of CD sensing, allowing for general schemes including additional ancillary modes.
We show that the ultimate quantum limit can be achieved by various optimal schemes. It includes not only Fock state input in direct sensing configuration but also twin-beam input in ancilla-assisted sensing configuration, for both of which photon number resolving detection needs to be performed as the optimal measurement setting. 
These optimal schemes offer a significant quantum enhancement even in the presence of additional system loss. 
The optimality of a practical scheme using a twin-beam state in direct sensing configuration is also investigated in details as a nearly optimal scheme for CD sensing when the actual CD signal is very small.
Alternative schemes involving single-photon sources and detectors are also proposed. 
This work paves the way for further investigations of quantum metrological techniques in chirality sensing.
\end{abstract}


\maketitle

\section{Introduction}
Measuring the optical response of media that consist of either chiral molecules~\cite{Greenfield2006, Micsonai2015} or chiral nanophotonic structures~\cite{Gansel2009, Passaseo2017, Collins2017} is of great importance in various scientific fields, from fundamentals to applications~\cite{Valev2013, Luo2017}. The chiral properties of a medium or a structure cause an asymmetric optical response upon illumination with either left- (LCP, or L) or right-handed circularly polarized light (RCP, or R). The optical response can be explained by an electric-magnetic coupling in the induced optical response, leading to effects such as circular dichroism (CD) and optical rotation~\cite{Bai2007, Oh2015}. While the former expresses the difference in absorption between LCP and RCP, the latter expresses a different phase accumulation upon propagation, leading to the rotation of the plane of linear polarization of light beam.

Particularly, the measurement of the CD signal has been widely used in various fields over the last few decades due to the simplicity of the measurement scheme combined with the rich information contained in the CD signal~\cite{Greenfield2006, Micsonai2015}. From the outcomes of CD measurement, relevant sample parameters under study may be estimated and we call this process CD sensing.
However, despite the great importance of CD measurement, the CD signal is usually very weak ($\sim10^{-3}$ to~$10^{-5}$ relative absorbance for chiral molecules) in realistic scenarios~\cite{Anson1974}. It is a nonlocal optical effect of the lowest order and only happens for molecules with broken inversion symmetry. Since the spatial extent of most molecules with respect to the incident field is negligibly small, the overall effect is rather tiny. When measuring it, one often struggles against the noise level, just similar to the case of gravitational wave detectors~\cite{Caves1981,Schnabel2010}. This limits the usefulness of CD spectroscopy to cases
where molecules are present either in high concentrations or in large volumes~\cite{Rodger1997, Kelly2005}, so that it is possible to accumulate enough signal. 

An obvious solution to the problem would be to increase the intensity of light that is incident on the analyte. However, this is not always an option due to optical damage that may occur in some situations~\cite{Neuman1999, Peterman2003, Taylor2015, Taylor2016}. Hence, one needs to look for alternative means to improve the sensing performance while keeping the incident power in the low-intensity regime.  
Also, for a fixed light source that is used in the measurements, the signal level in the CD measurements can be enhanced by using supporting photonic nanostructures~\cite{Hentschel2012, Valev2013, Wu2014, Yoo2015, Nesterov2016, Mohammadi2018, Garcia-Guirado2018, Vazquez-Guardado2018, Solomon2019, Graf2019, Droulias2020} or optical cavities~\cite{Feis2020}. 

A fundamentally different approach would be to use quantum states of light for sensing chiral properties of molecules. Quantum sensing schemes, in general, can reduce the noise below the shot-noise limit and consequently improve the signal-to-noise-ratio. For example, optical activity and optical rotatory dispersion of sucrose solution have been measured using single photons~\cite{Yoon2020} and polarization-entangled states~\cite{Tischler2016}, respectively. Both experimental studies clearly demonstrated the quantum enhancement in the estimation precision, i.e., sub-shot-noise limited sensing performance has been observed. Although schemes using quantum light emerge as a tool for ultimate sensing technology from diverse perspectives~\cite{Degen2017,Pirandola2018,Spedalieri2020}, sub-shot-noise limited quantum schemes for CD sensing have not yet been studied. 

In this work, we identify and investigate optimal CD sensing schemes that exploit quantum states of light consistent with any given energy constraint. For generality, we allow for ancilla-assisted sensing schemes, where entanglement between signal modes (i.e., LCP and RCP modes) and ancillary modes can play a role. To assess the CD sensing performance of various schemes in a comparable manner, we use multi-parameter estimation theory. The lower bound to the estimation uncertainty is defined using quantum Fisher information matrix (QFIM) and is called quantum Cram{\'e}r-Rao (QCR) bound. 
This allows to set the classical benchmark (CB) in CD sensing with a coherent state of light provided the optimal measurement is chosen. We then derive the ultimate quantum limit (UQL) to QCR bound that requires both the optimal quantum input state and the optimal measurement. 
It is shown that  even in realistic situations with additional system loss, the UQL always exhibits quantum enhancement in comparison with the CB. 
We show that the UQL can be achieved using Fock state input with photon number resolving detection (PNRD), for which ancillary modes are unnecessary. It is shown that using twin-beams as an input can also achieve the UQL in ancilla-assisted scheme, for which PNRD needs to be performed in both the signal and ancillary modes. Interestingly, the twin-beam state input is shown to be advantageous even in a direct sensing scheme that analyzes only the signal modes, in which no ancillary modes are used. The latter scheme provides a practical setting that achieves nearly ultimate QCR bound when losses are balanced at a moderate level and the difference in absorption between LCP and RCP modes is very small. Note that such a case applies to most CD sensing scenarios. 

\section{Theoretical Modelling}

\subsection{Circular dichroism sensing}
Illuminating a chiral medium with either LCP or RCP light results in transmission ($T$), reflection ($R$), and absorption ($A$) into the individual polarization modes. The intensity ratios are denoted by~$T_{jk}$,~$R_{jk}$, and~$A_{k}$ for~$j,k\in\{\text{L},\text{R}\}$, with the constraint~$\sum_{j} (T_{jk}+R_{jk})+A_{k}=1$, where the subscript~$k$~($j$) denotes the input (output) polarization. Apart from absorbance CD that can be quantified by the differential absorption, i.e.,~$A_\text{L}- A_\text{R}$, various alternative quantities can be measured to quantify the CD. A typical example would be transmission CD (TCD) defined as~$T_\text{LL}-T_\text{RR}$~\cite{Plum2008} or reflection CD defined as~$R_\text{LL}- R_\text{RR}$~\cite{Plum2016}. A polarization conversion in transmission or reflection may also occur, i.e.,~$T_{jk}\neq 0$ and~$R_{jk}\neq 0$ for~$j\neq k$, when the three-fold rotational symmetry does not hold with respect to the direction of incidence~\cite{Menzel2010}. It finally causes circular conversion dichroism, i.e.,~$T_\text{LR}\neq T_\text{RL}$~\cite{Schwanecke2008}.

\begin{figure}[t]
\centering
\includegraphics[width=0.48\textwidth]{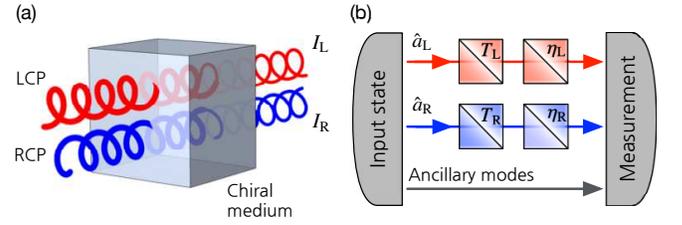}
\caption{(a) TCD is experimentally obtained by measuring the intensity difference of the transmitted LCP ($I_\text{L}$) and RCP ($I_\text{R}$) upon propagation through a chiral medium. (b) The ancilla-assisted CD sensing scheme is modeled quantum mechanically by the two signal modes corresponding to LCP and RCP and arbitrary ancillary modes that may be entangled with the signal modes. Beam splitters with transmittances~$T_\text{L (R)}$ and~$\eta_\text{L (R)}$ express the impact of the measurement device in each mode:~$T_j$ addresses the transmittance of each polarization mode through a chiral medium, whereas~$\eta_j$ addresses extra loss of each mode such as non-unity channel transmission and detection efficiency of a detector.}
\label{fig:model}
\end{figure}

In this work, not just for practical relevance with respect to realistic molecular samples or metamaterials that are typically considered, but also to eliminate the linear birefringence leading to unwanted polarization conversion, we focus on chiral media that preserve the four-fold rotational symmetry, for which~$T_{jk}= 0$ for~$j\neq k$ and~$R_{jj}= 0$ for all~$j$~\cite{Kwon2008, Plum2009a,Saba2013, Khan2019}. When illuminating media with a four-fold symmetry at normal incidence, the reciprocity further imposes~$R_\text{LR}=R_\text{RL}$~\cite{Bai2007, Kaschke2014}. Consequently, we have~$T_\text{LL}-T_\text{RR}=A_\text{R}-A_\text{L}$, and the intensity difference of the transmitted LCP ($I_\text{L}$) and RCP ($I_\text{R}$) becomes the key quantity of interest to be measured in usual CD measurement as illustrated in Fig.~\ref{fig:model}(a). Note that this is not a restrictive scenario but generally valid in usual scenarios when chiral molecules in solution are randomly oriented relative to the incident field. Considering a more general type of measurement that does not directly yield a parameter value under study, we use an estimator to estimate the quantity of TCD, defined as~$\Gamma_{-}\equiv T_\text{L}-T_\text{R}$, where~$T_{j}\equiv T_{jj}$. 

For the quantum mechanical description of CD or TCD sensing, let us consider, for generality, an ancilla-assisted scheme as shown in Fig.~\ref{fig:model}(b). 
The scheme consists of the two signal modes~$\hat{a}_\text{L}$ and~$\hat{a}_\text{R}$ that correspond to LCP and RCP modes, respectively and arbitrary number of ancillary modes. Such a general setup allows to consider correlated input states among the signal modes and ancillary modes, when necessary.
The transmission of each signal mode is described by a beam splitter with transmittance~$T_\text{L (R)}$. An extra loss that occurs outside an analyte (e.g., non-unity channel transmission or detection efficiency) can also be described by another beam splitter with transmittance~$\eta_\text{L (R)}$~\cite{Loudonbook}. For calculation of the output state, the two consecutive beam splitters in each signal mode can be treated as a single beam splitter, but the transmittances~$T_j$ and~$\eta_j$ have to be kept separate because only the parameters~$T_j$ are of interest in sensing while the factors~$\eta_j$ degrade the sensing performance. We also assume that the ancilla modes are lossless, which can be held in a controlled manner in many scenarios.
The associated input-output relation for the signal mode~$j\in \{\text{L},\text{R}\}$ is written as
\begin{align}
\hat{a}_{j}\rightarrow \sqrt{T_{j}\eta_{j}}\hat{a}_{j} + \sqrt{\eta_{j}(1-T_{j})}\hat{b}_{j} + \sqrt{(1-\eta_{j})(1-T_{j})}\hat{c}_{j},
\label{eq:Input_output_relation}
\end{align}
where~$\hat{b}_{j}$ and~$\hat{c}_{j}$ are virtual input modes associated with the chiral medium and the system loss respectively. Equation~\eqref{eq:Input_output_relation} is applied to the two signal modes of the total input state~$\vert \Psi_\text{in} \rangle$ containing ancillary modes. The resultant output state~$\hat{\rho}_\text{out}$ is measured using a chosen quantum measurement, yielding the outcomes~$\boldsymbol{m}$. From these, the TCD parameter~$\Gamma_{-}$ is estimated. This is the general CD sensing scheme we aim to investigate in this work. 

\subsection{Quantum multiparameter estimation theory}
The precision of CD sensing, a figure of merit which we consider in this work, can be formulated via quantum multi-parameter estimation theory~\cite{Szczykulska2016, Liu2020}. Consider an arbitrary pure state~$\vert \Psi_\text{in} \rangle$ as an input and suppose that the two transmittance parameters,~${\boldsymbol{T}}=(T_\text{L},T_\text{R})^\text{T}$, shall be estimated by an unbiased estimator from the measurement results~${\boldsymbol{m}}$ that have been drawn from a conditional probability~$p({\boldsymbol{m}}\vert {\boldsymbol{T}})$. In this case, one can find that the~$2\times 2$ covariance matrix~$\text{Cov}({\boldsymbol{T}})=\langle ({\boldsymbol{T}}-\langle {\boldsymbol{T}}\rangle)({\boldsymbol{T}}-\langle {\boldsymbol{T}}\rangle)^\text{T}\rangle$ obeys
\begin{align}
\text{Cov}({\boldsymbol{T}}) \ge \frac{{\boldsymbol{F}}^{-1}}{\nu},
\label{eq:CR_Inequality}
\end{align}
where~$\nu$ is the number of measurements being repeated and~${\boldsymbol{F}}$ is the Fisher information matrix (FIM) defined as~\cite{Helstrom1976,Paris2009}
\begin{align}
{\boldsymbol{F}}=\begin{pmatrix} F_\text{LL} & F_\text{LR} \\ F_\text{RL} & F_\text{RR}  \end{pmatrix},
\end{align}
where the matrix elements are written as
\begin{align}
F_{jk}=\sum_{{\boldsymbol{m}}} \frac{1}{p({\boldsymbol{m}}\vert {\boldsymbol{T}})} \frac{\partial p({\boldsymbol{m}}\vert {\boldsymbol{T}})}{\partial T_{j} }\frac{\partial p({\boldsymbol{m}}\vert {\boldsymbol{T}})}{\partial T_{k} },
\label{eq:FIM}
\end{align}
where~$j,k\in \{ R, L\}$. The lower bound in Eq.~\eqref{eq:CR_Inequality} is called Cram{\' e}r-Rao (CR) bound and can always be saturated by a maximum likelihood method in the limit of large~$\nu$~\cite{Braunstein1992}. The CR bound can be further reduced via optimization of a measurement setting, leading to~\cite{Helstrom1976,Paris2009}
\begin{align}
\text{Cov}({\boldsymbol{T}}) \ge \frac{{\boldsymbol{F}}^{-1}}{\nu}\ge \frac{{\boldsymbol{H}}^{-1}}{\nu},
\label{eq:QCR_Inequality}
\end{align}
where~${\boldsymbol{H}}$ denotes the QFIM defined by
\begin{align}
H_{jk}=\text{Tr}\left[\hat{\rho}_{\boldsymbol{T}} \frac{\hat{\cal L}_j \hat{\cal L}_k+\hat{\cal L}_k \hat{\cal L}_j}{2}\right],
\label{eq:QFIM}
\end{align} 
with~$\hat{\cal L}_j$ being a symmetric logarithmic derivative (SLD) operator associated with mode~$j$~\cite{Braunstein1994}. It is a solution of the equation
\begin{align}
\frac{\partial \hat{\rho}_{\boldsymbol{T}}}{\partial T_j}=\frac{1}{2}\left(\hat{\rho}_{\boldsymbol{T}}\hat{\cal L}_j+\hat{\cal L}_j \hat{\rho}_{\boldsymbol{T}}\right)
\label{eq:SLD_def}
\end{align}
for the parameter-encoded output state~$\hat{\rho}_{\boldsymbol{T}}$. 
Here,~${\boldsymbol{F}}^{-1}$ and~${\boldsymbol{H}}^{-1}$ are understood as the inverse on their support if the matrices are singular, i.e., not invertible~\cite{Ge2018}.
Since the independent SLD operators~$\hat{\cal L}_\text{L}$ and~$\hat{\cal L}_\text{R}$ commute, the optimal measurement setting can be constructed over the common eigenbasis of the commuting SLD operators~\cite{Baumgratz2016}. Thus, the lower bound in Eq.~\eqref{eq:QCR_Inequality}, called QCR bound, is saturable in CD sensing~\cite{Matsumoto2002}. 

Decomposing the state into the diagonalized bases, i.e.,~$\hat{\rho}_{\boldsymbol{T}}=\sum_n p_n\ket{\psi_n}\bra{\psi_n}$ with~$\langle \psi_n\vert\psi_m\rangle=\delta_{n,m}$, one can write the SLD operator as
\begin{align}
\hat{\cal L}_j
&=\sum_{n}\frac{\partial_j p_n}{p_n}\ket{\psi_n}\bra{\psi_n}+2\sum_{n\neq m}\frac{p_n-p_m}{p_n+p_m}\langle \psi_m\vert \partial_j \psi_n\rangle \ket{\psi_m}\bra{\psi_n},
\label{eq:SLD_decomposed}
\end{align}
where summation runs over~$n, m$ for which~$p_n+p_m\neq0$ and~$\partial_j\equiv \partial/\partial T_j$ for~$j\in\{\text{L},\text{R}\}$. Particularly when~$\vert \partial_j \psi_n\rangle=0$, the SDL operator~$\hat{\cal L}_j$ of Eq.~\eqref{eq:SLD_decomposed} becomes~$\hat{\cal L}_j=\sum_{n} (p_n)^{-1} (\partial_j p_n) \ket{\psi_n}\bra{\psi_n}$, for which the bases~$\{\ket{\psi_n}\bra{\psi_n}\}$ constitute the set of the optimal measurement bases~\cite{Paris2009}.

An alternative way to find the QFIM is to use the relation between Bures distance~${\cal D}_\text{B}^2$~\cite{Bures1969, Braunstein1994,Facchi2010}, quantum fidelity~${\cal F}$~\cite{Uhlmann1976,Jozsa1994}, and QFIM. In our case, the QFIM~$H_{jk}$ is related to the Bures distance~${\cal D}_\text{B}^{2}$ for the infinitesimally close states~$\hat{\rho}_{\boldsymbol{T}}$ and~$\hat{\rho}_{\boldsymbol{T}+\text{d}\boldsymbol{T}}$. It can be written as~\cite{Liu2020}
\begin{align}
\sum_{j,k\in\{\text{L},\text{R}\}} H_{jk} \text{d}T_j\text{d}T_k = 4 {\cal D}_\text{B}^{2}(\hat{\rho}_{\boldsymbol{T}}, \hat{\rho}_{\boldsymbol{T}+\text{d}\boldsymbol{T}}),
\end{align} 
where the Bures distance can be written in terms of quantum fidelity as 
\begin{align}
{\cal D}_\text{B}^{2}(\hat{\rho}_{\boldsymbol{T}}, \hat{\rho}_{\boldsymbol{T}+\text{d}\boldsymbol{T}}) = 2\left[ 1-\sqrt{{\cal F}(\hat{\rho}_{\boldsymbol{T}}, \hat{\rho}_{\boldsymbol{T}+\text{d}\boldsymbol{T}})}\right]
\end{align}
and the quantum fidelity is defined as
\begin{align}
{\cal F}(\hat{\rho}_{\boldsymbol{T}}, \hat{\rho}_{\boldsymbol{T}+\text{d}\boldsymbol{T}}) = \left(\text{Tr}\sqrt{\sqrt{\hat{\rho}_{\boldsymbol{T}}}\hat{\rho}_{\boldsymbol{T}+\text{d}\boldsymbol{T}}\sqrt{\hat{\rho}_{\boldsymbol{T}}}}\right)^2.
\label{eq:quantum_fidelity}
\end{align}
Thus, the calculation of quantum fidelity leads to the calculation of QFIM. 

The matrix inequality of Eq.~\eqref{eq:QCR_Inequality} reads 
\begin{align}
{\boldsymbol{n}}^\text{T}\text{Cov}(\boldsymbol{T}){\boldsymbol{n}} \ge \frac{{\boldsymbol{n}}^\text{T}{\boldsymbol{F}}^{-1}{\boldsymbol{n}}}{\nu}\ge \frac{{\boldsymbol{n}}^\text{T}{\boldsymbol{H}}^{-1}{\boldsymbol{n}}}{\nu},
\label{eq:Rev_QCR_Inequality}
\end{align}
for an arbitrary two-dimensional real vector~${\boldsymbol{n}}$~\cite{Pezze2017}. This applies when a global parameter~$\Gamma=\sum_{j}n_j T_j$, defined as a linear combination of multiple parameters, is estimated~\cite{Rubio2020, Knott2016, Guo2020, Oh2020, Gross2020}. For CD sensing,~$\Gamma_{-}=\boldsymbol{n}^{\text{T}}\boldsymbol{T}$ with~$\boldsymbol{n}=(1,-1)$, so we write~${\boldsymbol{n}}^\text{T}\text{Cov}(\boldsymbol{T}){\boldsymbol{n}}\equiv\text{Var}(\Gamma_{-})$. From now on, let us drop~$\nu$ as it appears everywhere. 

In the next sections, we use the QCR bound to investigate the lower bounds to the estimation uncertainty or equivalently, the precision of CD sensing for various input states of light. Individual cases are compared with the UQL which we shall derive below. One can see then what kinds of quantum states can achieve the UQL with and without assistance of ancillary modes. Furthermore, the QCR bounds and the UQL are also compared with the CR bounds for a particular measurement setting we choose depending on the input state considered. 
this constitutes an explicit specification of the measurement achieving the UQL. 

\section{Quantum Cram{\'e}r-Rao bound}

\subsection{Classical benchmark}
To derive for referential purposes the optimal QCR bound that is obtainable by using only classical light, let us consider a product of coherent states as an input state in Fig.~\ref{fig:model}(b), i.e.,~$\vert \alpha_\text{L} \rangle \vert \alpha_\text{R} \rangle =\hat{D}_\text{L}(\alpha_\text{L})\hat{D}_\text{R}(\alpha_\text{R})\vert 0\rangle \vert 0\rangle$ in a direct sensing configuration for simplicity, but without loss of generality. The coherent states are characterized by the average photon number~$N_j=\vert \alpha_j\vert^2$ and the displacement operators are represented by~$\hat{D}_j(\alpha_j)=\exp[\alpha \hat{a}_j^{\dagger}-\alpha^{*}\hat{a}_j]$. 
Applying the input-output relation of Eq.~\eqref{eq:Input_output_relation}, the output state can be written as
\begin{align}
\vert \Psi_\text{out} \rangle_\text{coh}=\vert \alpha_\text{L}^\text{(out)} \rangle \vert \alpha_\text{R}^\text{(out)} \rangle
\label{eq:Output_coherent}
\end{align}
with~$\alpha_j^\text{(out)}=\sqrt{\eta_j T_j} \alpha_j$. For such a pure output state, the QFIM of Eq.~\eqref{eq:QFIM} can be calculated via~\cite{Paris2009, Knott2016, Baumgratz2016}
\begin{align}
H_{jk} 
&=\frac{1}{2}\langle \Psi_\text{out}\vert \left( \hat{\cal L}_j \hat{\cal L}_k +\hat{\cal L}_k \hat{\cal L}_j \right)\vert \Psi_\text{out}\rangle,
\label{eq:QFIM_Pure}
\end{align}
where the SLD operator~$\hat{\cal L}_j$ can be written for a pure state~$\vert \Psi_\text{out}\rangle$ as
\begin{align}
\hat{\cal L}_j 
= 2\partial_j \vert \Psi_\text{out} \rangle \langle \Psi_\text{out} \vert.
\label{eq:SLD_Pure}
\end{align}
\begin{figure}[b]
\centering
\includegraphics[width=0.48\textwidth]{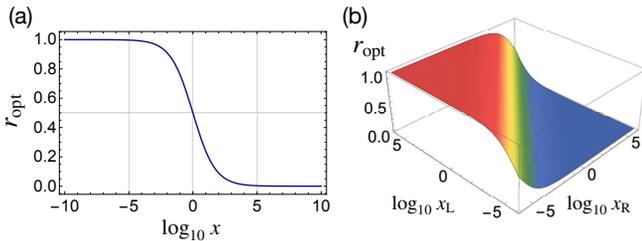}
\caption{(a) The optimal ratio~$r_\text{opt}$ as a function of logarithmic~$x=\eta_\text{L} T_\text{R}  / \eta_\text{R} T_\text{L}~$ for a coherent state input. The~$x$ is replaced by~$x= \eta_\text{L}T_\text{R} (1-\eta_\text{R} T_\text{R})  / \eta_\text{R} T_\text{L}(1-\eta_\text{L}T_\text{L})$ for the optimal state input achieving the UQL to the precision of CD sensing.
(b) The optimal ratio~$r_\text{opt}$ is shown as a function of logarithmic~$x_\text{L}=T_\text{L}$ and~$x_\text{R}=T_\text{R}$ for a coherent state input when~$\eta_\text{L}=\eta_\text{R}$. The axes labels are transformed to~$x_\text{L}=T_\text{L}(1- \eta_\text{L}T_\text{L})$ and~$x_\text{R}=T_\text{R}(1-\eta_\text{R}T_\text{R})$ for the optimal state input. 
}
\label{fig:ratio_opt}
\end{figure}

Through some algebraic calculation (see Appendix~\ref{appendix:QFIM_coherent} for details), one can find that the QFIM for~$\boldsymbol{T}$ with a coherent state input is diagonalized and written as
\begin{align}
\boldsymbol{H}_\text{coh} = \text{diag}\left( \frac{\eta_\text{L} N_\text{L} }{T_\text{L}}, \frac{\eta_\text{R} N_\text{R}}{T_\text{R}}\right).
\label{eq:QFIM_coh}
\end{align}
It is clear that~$\boldsymbol{H}_\text{coh} \rightarrow \boldsymbol{0}$ as~$\eta_\text{L/R}\rightarrow0$.
By substituting~$\boldsymbol{H}_\text{coh}$ to Eq.~\eqref{eq:Rev_QCR_Inequality}, the QCR bound to the estimation uncertainty of~$\Gamma_{-}$ can thus be written as
\begin{equation}
\text{Var}(\Gamma_{-})_\text{coh} = \frac{T_\text{L}}{\eta_\text{L} N_{\text{L}}} + \frac{T_\text{R}}{\eta_\text{R} N_{\text{R}}}.
\label{eq:QCR_coh}
\end{equation}
Defining the ratio~$r = N_\text{L}/ N_\text{tot}$ for the total average intensity in the signal modes~$N_\text{tot}=N_\text{L}+N_\text{R}$, which we fix throughout this work as a constraint, we find that the optimal ratio can be written as
\begin{align}
r_\text{opt} = \frac{1}{1 + \sqrt{\frac{\eta_\text{L} T_\text{R}}{\eta_\text{R}T_\text{L}}}},
\label{eq:Optimal_ratio}
\end{align}
for which the QCR bound of Eq.~\eqref{eq:QCR_coh} is minimized and thus reads 
\begin{align}
\text{Var}(\Gamma_{-})_\text{coh}^\text{opt} = \frac{1}{N_\text{tot}}\left(\sqrt{\frac{T_\text{L}}{\eta_\text{L}}}+\sqrt{\frac{T_\text{R}}{\eta_\text{R}}}\right)^{2}.
\label{eq:QCR_coh_opt}
\end{align}
The optimal ratio~$r_\text{opt}~$ of Eq.~\eqref{eq:Optimal_ratio} is presented in Fig.~\ref{fig:ratio_opt}(a) as a function of~$\eta_\text{L}  T_\text{R} / \eta_\text{R} T_\text{L}~$ in log scale, while shown in Fig.~\ref{fig:ratio_opt}(b) as a function of~$T_\text{L}$ and~$T_\text{R}$ in log scale for balanced losses, i.e.,~$\eta_\text{L}=\eta_\text{R}$. They clearly show that more energy needs to be injected into a more lossy signal mode to keep the optimal intensity balance between the signal modes, written as
\begin{align}
N_\text{L}:N_\text{R}=\sqrt{\frac{T_\text{L}}{\eta_\text{L}}} : \sqrt{\frac{T_\text{R}}{\eta_\text{R}}}.
\end{align}
In most cases, the difference in transmission between LCP and RCP light is extremely small and in good approximation it can be assumed that they are close to equal, i.e.,~$T_j = T_k$. Provided losses are balanced~$\eta_j=\eta_k$, the same amount of energies, i.e.,~$N_\text{L}=N_\text{R}$, would be, to a very good approximation, the optimal choice in a classical sensing scheme, for which~$\text{Var}(\Gamma_{-})_\text{coh}^\text{opt} = 4 T/\eta N_\text{tot}$ with~$T\equiv T_\text{L/R}$ and~$\eta\equiv \eta_\text{L/R}$. In cases where the two transmittances cannot be assumed to be equal, our findings can be combined with an adaptive scheme~\cite{Valeri2020, Nolan2020}. There, the input energies between LCP and RCP modes are adjusted in real-time, based on a prior information about the parameter being updated over repetition of the measurement.

One could consider an ancilla-assisted configuration with classically correlated coherent state input that can be written as
\begin{align}
\hat{\rho}_\text{coh}=\int p(\alpha_\text{L}, \alpha_\text{R}, \alpha_\text{A}) \vert \alpha_\text{L}, \alpha_\text{R}, \alpha_\text{A}\rangle \langle \alpha_\text{L}, \alpha_\text{R}, \alpha_\text{A}\vert \text{d}\alpha_\text{L}\text{d}\alpha_\text{R}\text{d}\alpha_\text{A}
\label{eq:classical_correlation}
\end{align}
with~$\text{Tr}[\hat{a}^{\dagger}_{j}~\hat{a}_{j}\hat{\rho}_\text{coh}]=N_{j}$. Applying the convexity of the QFIM~\cite{Takeoka2017}, one can prove that the QCR bound to be obtained for the input of Eq.~\eqref{eq:classical_correlation} is always equal to or greater than the CB of Eq.~\eqref{eq:QCR_coh_opt}. Therefore, neither ancillary modes nor classical correlation are useful here.

\subsection{Ultimate quantum limit}\label{sec:UQL}
Let us now derive the UQL on the estimation uncertainty of the TCD parameter~$\Gamma_{-}$. When~$\eta_\text{L/R}=1$, the maximum QFIM for two intensity parameters~$(T_\text{L}, T_\text{R})$, optimized over all input states, has been found in Ref.~\cite{Nair2018} and can be written as
\begin{align}
{\boldsymbol{H}}_\text{max}^\text{lossless}=\text{diag}\left(\frac{N_\text{L}}{T_\text{L}(1-T_\text{L})},\frac{N_\text{R}}{T_\text{R}(1-T_\text{R})}\right).
\label{eq:Max_QFIM_for_T_lossless}
\end{align}
It has been shown that the QCR bound associated with~${\boldsymbol{H}}_\text{max}^\text{lossless}$ can be achieved in general by so-called number-diagonal signal states in ancilla-assisted scheme~\cite{Nair2018} or by Fock state input without ancilla modes when~$N_\text{L}$ and~$N_\text{R}$ are integers~\cite{Adesso2009}.

In the presence of loss (i.e.,~$\eta_\text{L/R}\neq 1$), the SLD operators~$\hat{\cal L}_{j}$ for Eq.~\eqref{eq:Max_QFIM_for_T_lossless} is modified to~$\hat{\cal S}_{j}$, written as (see Appendix~\ref{appendix:SLD} for details)
\begin{align}
\hat{\cal S}_j= \eta_j \hat{\cal L}_j,
\end{align}
for~$j\in\{\text{L},\text{R}\}$. This leads~${\boldsymbol{H}}_\text{max}^\text{lossless}$ of Eq.~\eqref{eq:Max_QFIM_for_T_lossless} to be written as
\begin{align}
{\boldsymbol{H}}_\text{max}=\text{diag}\left(\frac{\eta_\text{L} N_\text{L}}{T_\text{L}(1-\eta_\text{L}T_\text{L})},\frac{\eta_\text{R}N_\text{R}}{T_\text{R}(1-\eta_\text{R}T_\text{R})}\right).
\label{eq:Max_QFIM_for_T_lossy}
\end{align}
This is the maximum QFIM for two intensity parameters~$(T_\text{L}, T_\text{R})$ in the presence of loss. It is clear that~${\boldsymbol{H}}_\text{max} \rightarrow \boldsymbol{0}$ as~$\eta_\text{L/R}\rightarrow0$.

The UQL to the estimation uncertainty~$\text{Var}(\Gamma_{-})$ can be readily obtained by substituting Eq.~\eqref{eq:Max_QFIM_for_T_lossy} into Eq.~\eqref{eq:Rev_QCR_Inequality}, resulting in
\begin{align}
\text{Var}(\Gamma_{-})_\text{UQL} = \frac{T_\text{L} (1 - \eta_\text{L} T_\text{L})}{\eta_\text{L} N_\text{L}} +  \frac{T_\text{R} (1 - \eta_\text{R} T_\text{R})}{\eta_\text{R} N_\text{R}}.
\label{eq:QCR_UQL}
\end{align}
This is the UQL to the estimation uncertainty or equivalently the precision of CD sensing for arbitrarily given~$N_\text{L}$ and~$N_\text{R}$.
The SLD operator~$\hat{\cal L}_{-}$ for the parameter~$\Gamma_{-}$ is obtained by using Eq.~\eqref{eq:SLD_def} and
\begin{align}
\frac{\partial \hat{\rho}_{\boldsymbol{T}}}{\partial \Gamma_{-}} 
=\frac{1}{2}\left(\hat{\rho}_{\boldsymbol{T}}\hat{\cal L}_{-} + \hat{\cal L}_{-} \hat{\rho}_{\boldsymbol{T}}\right).
\end{align}
It can thus be shown to be
\begin{align}
\hat{\cal L}_{-} = \frac{1}{2}\left(\hat{\cal S}_\text{L} - \hat{\cal S}_\text{R}\right).
\end{align}
One can easily show that the optimal ratio~$r_\text{opt}$ that minimizes~$\text{Var}(\Gamma_{-})_\text{UQL}$ of Eq.~\eqref{eq:QCR_UQL} can be written as
\begin{align}
r_\text{opt} = \frac{1}{1 + \sqrt{\frac{ \eta_\text{L} T_\text{R} (1- \eta_\text{R}T_\text{R} )}{\eta_\text{R}T_\text{L} (1-\eta_\text{L}T_\text{L}) }}},
\label{eq:Optimal_ratio_ultimate}
\end{align}
for which 
\begin{align}
\text{Var}(\Gamma_{-})_\text{UQL}^\text{opt} = \frac{1}{N_\text{tot}}\left(\sqrt{\frac{T_\text{L}(1- \eta_\text{L}T_\text{L})}{\eta_\text{L}}}+\sqrt{\frac{T_\text{R}(1-\eta_\text{R}T_\text{R})}{\eta_\text{R}}}\right)^{2}.
\label{eq:QCR_UQL_opt}
\end{align}
This is the UQL to the precision of CD sensing for the optimal raio between~$N_\text{L}$ and~$N_\text{R}$. It is obtained by the optimal input whose signal modes satisfying the optimal intensity ratio of Eq.~\eqref{eq:Optimal_ratio_ultimate} and the optimal measurement setting. The UQL applies to both cases with and without ancillary modes. Furthermore, the optimal schemes for scenarios without excess loss found in Ref.~\cite{Nair2018} can be used to reach the UQL of Eq.~\eqref{eq:QCR_UQL_opt} in lossy scenarios.
 
\begin{figure}[t]
\centering
\includegraphics[width=0.48\textwidth]{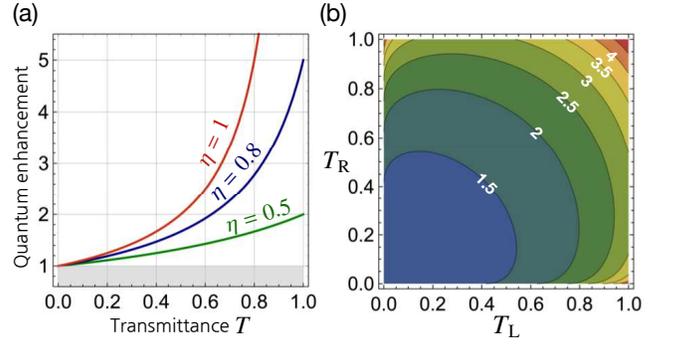}
\caption{(a) Quantum enhancement~$\text{Var}(\Gamma_{-})_\text{CB}/\text{Var}(\Gamma_{-})_\text{UQL}$ in terms of transmittance~$T$ for~$\eta=1, 0.8, 0.5$ when~$T_\text{L/R}=T$ and~$\eta_{L/R}=\eta$ can be assumed. 
(b) Quantum enhancement~$\text{Var}(\Gamma_{-})_\text{CB}/\text{Var}(\Gamma_{-})_\text{UQL}$ in terms of~$T_\text{L}$ and~$T_\text{R}$ for~$\eta=0.8$. 
}
\label{fig:QE}
\end{figure}

Comparing the UQL of Eq.~\eqref{eq:QCR_UQL_opt} with the CB of Eq.~\eqref{eq:QCR_coh_opt}, one can see that the quantum enhancement is achieved by the factors of~$(1-\eta_\text{L}T_\text{L})$ and~$(1-\eta_\text{R}T_\text{R})$ in the numerator of the respective terms, but diminishes with loss, i.e., as~$\eta_\text{L/R}\rightarrow 0$. Note that both QCR bounds of Eqs.~\eqref{eq:QCR_coh_opt} and \eqref{eq:QCR_UQL_opt} scale with~$N_\text{tot}$, i.e., following the shot-noise scaling in terms of the total energy~$N_\text{tot}$, as in the single loss parameter estimation case~\cite{Adesso2009}. The optimal ratio~$r_\text{opt}$ of Eq.~\eqref{eq:Optimal_ratio_ultimate} exhibits the same behavior as shown in Figs.~\ref{fig:ratio_opt}(a) and (b), but with~$\eta_\text{L}T_\text{R} (1-\eta_\text{R} T_\text{R}) / \eta_\text{R}T_\text{L} (1-\eta_\text{L}T_\text{L})~$ and~$x_j=T_j (1-\eta_j T_j)$ for~$j=\text{L},\text{R}$, respectively. The optimal ratio~$r_\text{opt}$ can also be understood as the optimal balance of the average intensities between the LCP and RCP modes, written as
\begin{align}
N_\text{L}:N_\text{R} = \sqrt{\frac{T_\text{L}(1-\eta_\text{L}T_\text{L})}{\eta_\text{L}}}: \sqrt{\frac{T_\text{R}(1-\eta_\text{R}T_\text{R})}{\eta_\text{R}}}.
\label{eq:Intensity_balance_UQL}
\end{align}
Again, in most cases,~$T_j \approx T_k$ and~$\eta_j \approx \eta_k$, so the same amount of energies, i.e.,~$N_\text{L}=N_\text{R}$, would be the optimal choice in the ultimate CD sensing scheme, for which~$\text{Var}(\Gamma_{-})_\text{UQL}^\text{opt} = 4 T(1-\eta T)/\eta N_\text{tot}$ with~$T\equiv T_\text{L/R}~$ and~$\eta\equiv \eta_\text{L/R}$. In this case, the quantum enhancement of the UQL as compared to the CB can be quantified by the ratio defined as 
\begin{align}
\frac{\text{Var}(\Gamma_{-})_\text{coh}^\text{opt}}{\text{Var}(\Gamma_{-})_\text{UQL}^\text{opt}}=\frac{1}{1-\eta T}.
\end{align}
Note that this enhancement factor diverges as~$\eta T \rightarrow 1$, so the infinite-fold enhancement can be in principle achieved or a huge quantum enhancement can be exploited in well-controlled situations. The enhancement is degraded as~$\eta$ decreases in lossy cases, e.g., the maximal enhancement is only two-fold for~$\eta=0.5$. The enhancement factor is presented in Fig.~\ref{fig:QE}(a) as a function of transmittance~$T$ for~$\eta=1, 0.8$, and~$0.5$. It clearly shows that the quantum enhancement is sensitive to the loss parameter~$\eta$, so reducing the loss in a sensing setup is crucial to increase the quantum enhancement for a given~$T_\text{L}\approx T_\text{R}=T$ in CD sensing. We, nevertheless, stress that the quantum enhancement factor is always greater than unity unless either~$\eta$ or~$T$ is zero. Figure~\ref{fig:QE}(b) shows an overall quantum enhancement in terms of arbitrary~$T_\text{L}$ and~$T_\text{R}$ for balanced loss~$\eta=0.8$ chosen as an example.

\subsection{Fock state input}

We now show that the Fock state input~$\vert N_\text{L}\rangle \vert N_\text{R}\rangle$ without using ancillary modes can achieve the UQL. Through the beam splitter transformation of Eq.~\eqref{eq:Input_output_relation}, the output state can be written as
\begin{align}
\hat{\rho}_\text{Fock}=\sum_{m_\text{L},m_\text{R}} p(m_\text{L},m_\text{R}\vert \boldsymbol{T}) \vert m_\text{L}, m_\text{R}\rangle \langle m_\text{L}, m_\text{R}\vert,
\label{eq:Output_Fock}
\end{align}
where 
\begin{align}
p(m_\text{L},m_\text{R}\vert \boldsymbol{T}) =
\prod_{j=\text{L},\text{R}} \binom{N_j}{m_j}(\eta_j T_j)^{m_j}(1-\eta_j T_j)^{N_j-m_j}.
\end{align}
With this, one can show that the QFIM is equal to~${\boldsymbol{H}}_\text{max}$ of Eq.~\eqref{eq:Max_QFIM_for_T_lossy}, finally achieving the UQL of Eq.~\eqref{eq:QCR_UQL_opt} when the photon numbers~$N_\text{L}$ and~$N_\text{R}$ follow the optimal ratio of Eq.~\eqref{eq:Intensity_balance_UQL}. Therefore, Fock state input~$\vert N_\text{L}\rangle \vert N_\text{R}\rangle$ is the optimal state to reach the UQL to the precision in CD sensing. The UQL is inversely proportional to the total average photon number~$N_\text{tot}$, so it is recommended to increase the total intensity of an input state while keeping the optimal ratio of Eq.~\eqref{eq:Intensity_balance_UQL}. However, large Fock states with~$N_j\gg 1$ cannot be readily generated with current technology~\cite{Varcoe2000, Bertet2002, Waks2006}. As shown in Ref.~\cite{Nair2018}, an alternative way is to use~$N_j$ single-photons~\cite{Lounis2005, Eisaman2011, Meyer-Scott2020}, which also leads to the UQL on the precision of CD sensing.

\subsection{Twin-beam input}
Another useful quantum source of light is the so-called twin-beam. They have widely been used in many applications including quantum imaging~\cite{Genovese2016}, quantum illumination~\cite{Lloyd2008, Tan2008, Lopaeva2013, Nair2020}, and quantum sensing~\cite{Meda2017} due to the strong photon number correlation~\cite{Jedrkiewicz04,Bondani07,Blanchet08,Perina12}. 
The twin-beam state can be generated from a spontaneous parametric down conversion process~\cite{Burnham1970, Heidmann1987, Schumaker1985} and is formally written as the two-mode squeezed vacuum (TMSV) state, ~$\vert \text{TMSV}\rangle = \hat{S}_2(\xi)\ket{00}$ with the two-mode squeezing operator~$\hat{S}_2(\xi)=\exp[\xi^{*}\hat{a}\hat{b}-\xi \hat{a}^{\dagger}\hat{b}^{\dagger}]$ for~$\xi=re^{i\theta}$ with~$\{r,\theta \} \in\mathbb{R}$. As shown below, such TMSV states or twin-beams can be used for CD sensing in two ways. 

First, let us consider CD sensing scheme using two TMSV states~$\vert \text{TMSV}\rangle \otimes \vert \text{TMSV}\rangle$ in an ancilla-assisted configuration. Let us assume that the respective signal modes of the TSMV states are sent to LCP and RCP mode, while their respective ancillary modes are held losslessly. Such a setting has been shown to achieve the QFIM of Eq.~\eqref{eq:Max_QFIM_for_T_lossless} for~$(T_\text{L}, T_\text{R})$ in the absence of additional loss~\cite{Nair2018}. The analysis in Section~\ref{sec:UQL} implies that the same setting can be used to achieve the UQL of Eq.~\eqref{eq:QCR_UQL_opt} when the average intensities of the signal states of the two TMSV states satisfy the optimal ratio of Eq.~\eqref{eq:Intensity_balance_UQL}. Therefore, the twin-beam input is the optimal state to reach the UQL to the precision in CD sensing in an ancilla-assisted configuration. One can find other optimal states in ancilla-assisted scheme according to the analysis in Ref.~\cite{Nair2018}.

A second way to use the TMSV state input is to inject the two modes of a single TMSV state into LCP and RCP modes, respectively. Note that ancillary modes are not considered in such direct sensing scheme and 
$N_\text{L}=N_\text{R}=\sinh^2 r\equiv N$ as an intrinsic feature of the twin-beam state.
The output state~$\hat{\rho}_{\boldsymbol{T}}$ can be obtained by using the input-output relation of Eq.~\eqref{eq:Input_output_relation} for the input state~$\vert \text{TMSV}\rangle$. In this particular case, analytical calculation of the QFIM using SLD operators is tricky, so we use the quantum fidelity of Eq.~\eqref{eq:quantum_fidelity} that can be calculated more easily using a closed expression~\cite{Marian2012, Banchi2015}. Leaving all the technical details to Appendix~\ref{appendix:fidelity}, we finally have the QFIM written as
\begin{align}
H_{jj} &=\frac{\chi_{j\bar{j}}\eta_j N}{T_j(1-\eta_j T_j)},
\label{eq:QFIM_TMSV_diagonal}\\
H_{jk} &=\frac{-\eta_j\eta_k N(N+1)}{1+\eta_j T_j (1-\eta_k T_k)N+\eta_k T_k(1-\eta_j T_j)N},\label{eq:QFIM_TMSV_off-diagonal}
\end{align}
with
\begin{align}
\chi_{j\bar{j}}=\frac{1-\eta_j T_j(1-\eta_{\bar{j}} T_{\bar{j}})+\eta_{\bar{j}} T_{\bar{j}}(1-\eta_j T_j)N}
{1+\eta_j T_j (1-\eta_{\bar{j}} T_{\bar{j}})N+\eta_{\bar{j}} T_{\bar{j}}(1-\eta_j T_j)N},
\end{align}
where~$j\neq k \in\{\text{L},\text{R}\}$,~$\bar{j} = \text{R}$ if~$j=\text{L}$, and vice versa.

In comparison with the QFIM of Eq.~\eqref{eq:Max_QFIM_for_T_lossy}, 
the diagonal element~$H_{jj}$ of Eq.~\eqref{eq:QFIM_TMSV_diagonal} contains the additional factor~$\chi_{jk}$ coming from the correlation between the signal modes. One can show that~$0 \le \chi_{jk}\le1$ holds, where the upper bound is reached when~$\eta_j T_j=0$ or~$\eta_k T_k=1$, while the lower bound is obtained when~$\eta_j T_j=1$ and~$\eta_k T_k=0$. The QCR bound for the estimation uncertainty~$\text{Var}(\Gamma_{-})$ can thus be written in terms of Eqs.~\eqref{eq:QFIM_TMSV_diagonal} and \eqref{eq:QFIM_TMSV_off-diagonal} as 
\begin{align}
\text{Var}(\Gamma_{-})_\text{TMSV}
= \frac{H_\text{LL}+H_\text{RR}+H_\text{LR}+H_\text{RL}}{H_\text{LL}H_\text{RR}-H_\text{LR}H_\text{RL}}.
\label{eq:QCR_TMSV}
\end{align}
It can be easily shown that the use of a TMSV state input in direct sensing scheme cannot achieve the UQL to the precision of CD sensing. However, one can find that the QCR bound~$\text{Var}(\Gamma_{-})_\text{TMSV}$ of Eq.~\eqref{eq:QCR_TMSV} becomes similar to the UQL at some regimes of parameters, which we elaborate on in more details below. 


\begin{figure}[b]
\centering
\includegraphics[width=0.48\textwidth]{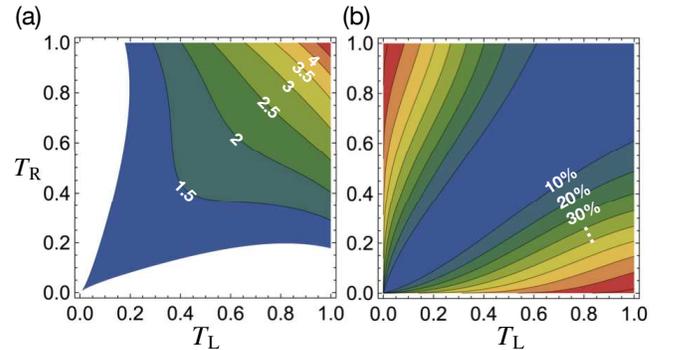}
\caption{(a) Quantum enhancement~$\text{Var}(\Gamma_{-})_\text{CB}/\text{Var}(\Gamma_{-})_\text{TMSV}$ in terms of~$T_\text{L}$ and~$T_\text{R}$ for balanced losses~$\eta=0.8$.
(b) The normalized difference between~$\text{Var}(\Gamma_{-})_\text{TMSV}$ and~$\text{Var}(\Gamma_{-})_\text{UQL}^\text{opt}$ for~$\eta=0.8$. Here, $N_\text{tot}=2$ is assumed as an example. 
}
\label{fig:TMSV}
\end{figure}

For comparison of~$\text{Var}(\Gamma_{-})_\text{TMSV}$ with the other cases, let us set~$N_\text{tot}=2N=2$ and~$\eta_\text{L}=\eta_\text{R}=0.8$ as example without loss of generality. In Fig.~\ref{fig:TMSV}(a), we show the quantum enhancement~$\text{Var}(\Gamma_{-})_\text{coh}^\text{opt}/\text{Var}(\Gamma_{-})_\text{TMSV}$ is limited to only the presented region. Such a beneficial region depends on the values of~$N$ and~$\eta_\text{L (R)}$, but in a particular region of interest for CD sensing, i.e., when~$T_\text{L}\approx T_\text{R}$, the enhancement is always present and significant. More interestingly and clearly, it can be shown that~$\text{Var}(\Gamma_{-})_\text{TMSV}= \text{Var}(\Gamma_{-})_\text{UQL}^\text{opt}$ holds up to the first-order in~$\delta T$ for~$T_\text{R}=T_\text{L}+\delta T$ when losses are equally balanced~$\eta_\text{L}=\eta_\text{R}$. Such a feature is evidently shown in Figs.~\ref{fig:TMSV}(a) and (b) around the region where~$T_\text{L}\approx T_\text{R}$. This indicates that in most cases when~$T_\text{L}$ and~$T_\text{R}$ can be assumed in good approximation as equal, one can use the direct sensing scheme with the TMSV state input as a practical scheme. The use of TMSV state input also promises quantum enhancement for any value of~$T\equiv T_\text{L}= T_\text{R}$, as already shown in Fig.~\ref{fig:QE}. This is an important finding as it opens a practical path towards exploiting of practical quantum resources in realistic CD sensing.

It is worth discussing the role of the average photon number~$N$ in~$\text{Var}(\Gamma_{-})_\text{TMSV}$. A noticeable behavior is revealed in the limit of large~$N$. Both~$\text{Var}(\Gamma_{-})_\text{UQL}^{\text{opt}}$ and~$\text{Var}(\Gamma_{-})_\text{coh}^{\text{opt}}$ approach zero as~$N\rightarrow \infty$, whereas~$\text{Var}(\Gamma_{-})_\text{TMSV}$ becomes
\begin{align}
\text{Var}(\Gamma_{-})_\text{TMSV}\vert_{N\rightarrow\infty} = \frac{(T_\text{L}-T_\text{R})^{2}(1-\eta_\text{L} T_\text{L})(1-\eta_\text{R} T_\text{R})}{1+\eta_\text{L} T_\text{L} (1-\eta_\text{R} T_\text{R})+\eta_\text{R} T_\text{R}(1-\eta_\text{L} T_\text{L})}.
\end{align}
This implies that the use of the TMSV state input outperforms the CB only when~$T_\text{L}\approx T_\text{R}$ or~$N$ is small. In other words, as~$N$ is reduced, the beneficial region in Fig.~\ref{fig:TMSV}(c) becomes wider, but never covers the entire region. This means that no quantum enhancement is obtained when either~$T_\text{L}$ or~$T_\text{R}$ is too small even in the limit~$N\rightarrow 0$. Such a region is of course not of much interest for CD sensing, but could be significant for other applications.

\subsection{Signal-to-noise ratio}
For an estimator of~$\hat{\Gamma}_{-}$, one can define the signal-to-noise ratio of an estimate~$\Gamma_{-}$ as 
\begin{align}
\text{SNR}=\frac{\langle \hat{\Gamma}_{-}\rangle^{2}}{\text{Var}( \Gamma_{-}) },
\end{align}
where~$\langle \hat{\Gamma}_{-}\rangle^{2}=\Gamma_{-}^{2}$ for an unbiased estimator.
Using Eq.~\eqref{eq:Rev_QCR_Inequality}, one can easily show that for a given input state the SNR is upper bounded as
\begin{align}
\text{SNR}\le \frac{\Gamma_{-}^{2}}{\text{Var}( \Gamma_{-}) _\text{QCR}},
\label{eq:SNR}
\end{align}
where~$\text{Var}( \Gamma_{-}) _\text{QCR}$ is the QCR bound to~$\text{Var}( \Gamma_{-})$. 
This shows that the upper bound of SNR becomes higher by increasing~$\Gamma_{-}$ while decreasing~$\text{Var}( \Gamma_{-}) _\text{QCR}$. 
In other words, precise sensing with small~$\text{Var}( \Gamma_{-}) _\text{QCR}$ yields high SNR, but its inverse does not hold. This implies that assessment of CD sensing in terms of SNR does not guarantee precise estimation of CD or TCD parameter. 
The equality in Eq.~\eqref{eq:SNR} can be saturated when the optimal measurement setting and the optimal estimator are used for a given state. A similar SNR inequality for a single parameter estimation has also been discussed in Ref.~\cite{Paris2009}.

\section{Measurements achieving the ultimate quantum limit}
Let us now consider particular measurement settings to examine if the CR bound reaches the QCR bound for individual cases. In this work we employ direct detection measurements at each output port in Fig.~\ref{fig:model}(b), which measures the intensities of the transmitted signal modes through a chiral medium and ancillary modes having been kept unaltered. In particular, a PNRD measurement yields the multi-dimensional photon number distribution for the measurement outcomes~$\boldsymbol{m}$ drawn from the underlying conditional probability~$p(\boldsymbol{m}\vert \boldsymbol{T})$. 

\subsection{Coherent state input}
For a coherent state input in a direct sensing configuration, the output state is given as Eq.~\eqref{eq:Output_coherent} and the probability distribution of detecting~$m_\text{L}$ and~$m_\text{R}$ photons at the respective output ports is written as
\begin{align}
p(m_\text{L},m_\text{R}\vert \boldsymbol{T})
= \prod_{j=\text{L},\text{R}}e^{-\eta_j T_j N_j} \frac{(\eta_j T_j N_j)^{m_j}}{m_j!}.
\label{eq:prob_PNRD_coherent}
\end{align}
Using Eq.~\eqref{eq:FIM}, one can show that the FIM with Eq.~\eqref{eq:prob_PNRD_coherent} is the same as QFIM of Eq.~\eqref{eq:QFIM_coh}, implying that the PNRD is the optimal measurement setting to reach the optimal classical bound of Eq.~\eqref{eq:QCR_coh} when~$N_\text{L}$ and~$N_\text{R}$ are arbitrary chosen or the CB of Eq.~\eqref{eq:QCR_coh_opt} when the optimal ratio between~$N_\text{L}$ and~$N_\text{R}$ is chosen. 

\subsection{Fock state input}
For a Fock state input without ancillary modes, the output state of Eq.~\eqref{eq:Output_Fock} is diagonalized over the photon number states~$\{ \vert m_\text{L}, m_\text{L}\rangle\}$. It is clear that the diagonalized basis is independent of the parameter~$\boldsymbol{T}$, so the second term in Eq.~\eqref{eq:SLD_decomposed} vanishes and consequently the FIM of Eq.~\eqref{eq:FIM} is the same as the QFIM of Eq.~\eqref{eq:Max_QFIM_for_T_lossy}. This indicates that the PNRD offers the optimal measurement setting for the case using Fock state inputs. The optimality of the PNRD can also be proved from the fact that the eigenstates of the corresponding SLD operator are the photon number states~$\{ \vert m_\text{L}, m_\text{L}\rangle\}$~\cite{Paris2009, Oh2019}. 

When multiple single photons are used~\cite{Adesso2009} instead of large Fock states that are yet unavailable with current technology~\cite{Lounis2005, Eisaman2011, Meyer-Scott2020}, we can use, instead of PNRD, single-photon detectors which are a well-established technology~\cite{Holzman2019}. This achieves the UQL. 

\subsection{Twin-beam input}
When using twin-beams in ancilla-assisted scheme, the QFIM of Eq.~\eqref{eq:Max_QFIM_for_T_lossless} has been shown to be achievable by performing PNRD in all the four modes, i.e., two signal and two ancillary modes~\cite{Nair2018}. As explained previously, such optimality of the measurement scheme also carries over to the measurement of CD in the presence of loss, consequently achieving the UQL.

To reach the same bound, one can use~$M=N/n$ copies of weakly squeezed TMSVs with the average photon number of~$n\ll1$ on each mode and perform direct detection on each two-mode output state~\cite{Nair2018}. Apart from placing less demands on high squeezing required in the twin-beam, weak fields with~$n\ll 1$ allow us to perform, instead of PNRD, single-photon detection~\cite{Holzman2019}.

\begin{figure}[t]
\centering
\includegraphics[width=0.48\textwidth]{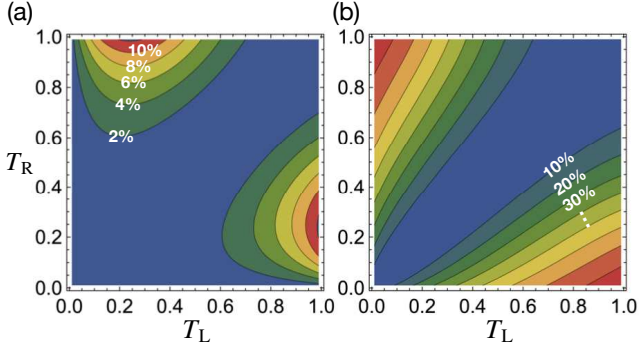}
\caption{
(a) The normalized difference between the CR bound for PNRD and the QCR bound~$\text{Var}(\Gamma_{-})_\text{TMSV}$ for balanced losses~$\eta=0.8$.
(b)The normalized difference between the CR bound for PNRD and the UQL~$\text{Var}(\Gamma_{-})_\text{UQL}^\text{opt}$ for~$\eta=0.8$. Here, $N_\text{tot}=2$ is assumed as an example.
}
\label{fig:TMSV_FI}
\end{figure}

For direct sensing scheme with a TMSV state input, the output state is a mixed state and not diagonalized over the photon number states. The photon number distribution of the output state is given as
\begin{align}
p(m_\text{L},m_\text{R}\vert \boldsymbol{T}) 
=\sum_{n=0}^{\infty}
\frac{N^{n}}{(N+1)^{n+1}}  \prod_{j=\text{L},\text{R}} f_{j}(n),
\end{align}
where~$f_{j}(n)=\binom{n}{m_j}(\eta_j T_j)^{m_j}(1-\eta_j T_j)^{n-m_j}$. In this case, we numerically calculate the~$\boldsymbol{F}$ of Eq.~\eqref{eq:FIM}, which gives rise to the CR bound. The latter is compared with the QCR bound~$\text{Var}(\Gamma_{-})_\text{TMSV}$ and the UQL~$\text{Var}(\Gamma_{-})_\text{UQL}$ for balanced losses~$\eta=0.8$ chosen as an example. They are shown in Figs.~\ref{fig:TMSV_FI}(a) and (b), respectively. The CR bound is not generally the same as the QCR bound~$\text{Var}(\Gamma_{-})_\text{TMSV}$, but they become extremely similar when~$T_\text{L}$ and~$T_\text{R}$ are close to each other, as shown in Fig.~\ref{fig:TMSV_FI}(a). This indicates that the CR bound can also be similar the UQL~$\text{Var}(\Gamma_{-})_\text{UQL}$ in the region where~$T_\text{L}\approx T_\text{R}$. The latter behavior is evident in Fig.~\ref{fig:TMSV_FI}(b). Especially, one can show that the CR bound becomes exactly the same as the other two bounds when~$T_\text{L}=T_\text{R}$. This means that one can use the direct sensing scheme with the twin-beam state input and PNRD as a practically optimal scheme for CD sensing when~$T_\text{L}\approx T_\text{R}$ can be assumed and losses are balanced. 

\section{Conclusion}
We have obtained the UQL on the precision of CD sensing and identified the optimal CD sensing schemes to achieve it. With the optimal schemes studied in this work, a significant quantum enhancement has been shown to be achievable even in the presence of loss. For most samples of chiral media being analyzed by CD measurement, the difference between the transmittance parameters~$T_\text{L}$ and~$T_\text{R}$ is very small. For such usual cases, we have proposed a practical CD sensing scheme to reach nearly the UQL, which requires only to use the twin-beam state as an input and to perform PNRD at the two signal modes.

The formalism we have used in this work can be immediately applied to linear dichroism sensing~\cite{Dorr1966, Rodger1997} or magnetic CD~\cite{Stephens1970, Chen1995}. The role of entanglement would be more significant when polarization conversion starts to be involved~\cite{Plum2008, Plum2016, Schwanecke2008, Menzel2010}, which was not considered in this work. The CD usually occurs in units of a single photon, which enabled us to model CD by linear beam splitters. However, it may occur in units of two photons, called two-photon CD (TPCD)~\cite{Tinoco1975, Power1975}. The latter needs to be modeled by non-linear beam splitters, where transmission takes place in units of two photons. It would be interesting to study optimal TPCD sensing schemes with quantum light. CD sensing with plasmonic chiral structures are often studied~\cite{Jeong2016}, for which the technique studied in this work can cooperate with the recently developed quantum plasmonic sensing techniques~\cite{Fan2015, Lee2016, Lee2020BookChapter}.
\\

\section*{acknowledgments}
C.L. thanks Xavier Garcia-Santiago for useful discussion. 
This work was  partially supported by the Deutsche Forschungsgemeinschaft (DFG, German Research Foundation) under Germany’s Excellence Strategy via the Excellence Cluster 3D Matter Made to Order (EXC-2082/1 – 390761711) and by the VIRTMAT project at KIT.
R.N. and M.G. are supported by the National Research Foundation Singapore (NRF-NRFF2016-02); NRF Singapore and L'Agence Nationale de la Recherche Joint Project (NRF2017-NRFANR004 VanQuTe); Singapore Ministry of Education (MOE2019-T1-002-015); FQXi (FQXi-RFP-IPW-1903).

\appendix
\section*{Appendix}

\section{QFIM for a coherent state input}\label{appendix:QFIM_coherent}
\setcounter{equation}{0}
\renewcommand{\theequation}{A\arabic{equation}}
\setcounter{figure}{0}
\renewcommand{\thefigure}{A\arabic{figure}}
The QFIM of Eq.~\eqref{eq:QFIM_Pure} can be rewritten as
\begin{align}
H_{jk} 
&= 4 \text{Re}\left( \langle \partial_j \Psi_\text{out}\vert \partial_k \Psi_\text{out}\rangle 
- \langle \partial_j \Psi_\text{out} \vert \Psi_\text{out}\rangle  \langle \Psi_\text{out} \vert \partial_k \Psi_\text{out}\rangle  \right).
\label{eq:QFIM_Pure_Rev}
\end{align}
For the output state of Eq.~\eqref{eq:Output_coherent}, the derivative is written by
\begin{align}
\vert \partial_j \Psi_\text{out}\rangle 
=\sqrt{\frac{\eta_j}{4T_j}}\left(\alpha_j \hat{a}_j^\dagger -\alpha_j^{*}\hat{a}_{j}\right)\vert \Psi_\text{out}\rangle,
\end{align}
causing that the second term of Eq.~\eqref{eq:QFIM_Pure_Rev} vanishes for all~$j, k$. The first term, on the other hand, is shown to be written as
\begin{align}
H_{jk}=\frac{\eta_j N_j}{T_j}\delta_{jk},
\end{align}
where $\delta_{jk}$ denotes the Kronecker delta. Thus, we have~$\boldsymbol{H}$ of Eq.~\eqref{eq:QFIM_coh} in the main text. 
\section{SLD operators}\label{appendix:SLD}
\setcounter{equation}{0}
\renewcommand{\theequation}{B\arabic{equation}}
\setcounter{figure}{0}
\renewcommand{\thefigure}{B\arabic{figure}}
The QFIM of Eq.~\eqref{eq:Max_QFIM_for_T_lossless} can be understood as the UQL to estimation of the total transmittance~${\cal T}_j$ of individual modes, for which the SLD operators are written as
\begin{align}
\frac{\partial \hat{\rho}}{\partial {\cal T}_j} = \frac{1}{2} \left(\hat{\rho} \hat{\cal L}_j + \hat{\cal L}_j \hat{\rho}\right).
\end{align}
Decomposing the total transmittance as~${\cal T}_j=\eta_j T_j$, one can find the SLD operators~$\hat{\cal S}_j$ for estimation of~$T_j$ as follows. 
\begin{align}
\frac{\partial \hat{\rho}}{\partial T_j} 
= \frac{\partial \hat{\rho}}{\partial {\cal T}_j}\frac{\partial {\cal T}_j}{\partial T_j} 
= \frac{\partial \hat{\rho}}{\partial {\cal T}_j}\eta_j 
= \frac{1}{2} \left(\hat{\rho} \hat{\cal S}_j + \hat{\cal S}_j \hat{\rho}\right),
\end{align}
where~$\hat{\cal S}_j= \eta_j \hat{\cal L}_j$. Therefore, we have the QFIM written as
\begin{align}
H_{jk}&=\frac{1}{2}\text{Tr}[\rho (\hat{\cal S}_j \hat{\cal S}_j + \hat{\cal S}_j \hat{\cal S}_j)]
=\frac{ \eta_j N_j }{ T_j (1-\eta_j T_j)}\delta_{jk},
\end{align}
where~$\delta_{jk}$ denotes the Kronecker delta. Thus, we have~${\boldsymbol{H}}_\text{max}$ of Eq.~\eqref{eq:Max_QFIM_for_T_lossy} in the main text and this is the modified UQL to estimation of $T_j$ in the presence of loss. 

\section{Quantum fidelity}\label{appendix:fidelity}
\setcounter{equation}{0}
\renewcommand{\theequation}{C\arabic{equation}}
\setcounter{figure}{0}
\renewcommand{\thefigure}{C\arabic{figure}}
For a TMSV state input, the output state~$\hat{\rho}_{\boldsymbol{T}}$ can be characterized by only the second-order moments, i.e., the covariance matrix~$\boldsymbol{V}$~\cite{Braunstein2005,Weedbrook2012}. Using the analytical form of quantum fidelity that has been found for covariance matrices~\cite{Marian2012, Banchi2015}, one can readily calculate the quantum fidelity for the TMSV state input. 

The covariance matrix~$\boldsymbol{V}$ is defined by~$V_{jk}=\text{Tr}[\hat{\rho}_{\boldsymbol{T}}\{ \hat{Q}_{j}-d_{j},\hat{Q}_{k}-d_{k}\}/2]$, where~$\{\hat{A},\hat{B}\}\equiv \hat{A}\hat{B}+\hat{B}\hat{A}$ and~$d_{j}=\text{Tr}[\hat{\rho}_{\boldsymbol{T}}\hat{Q}_{j}]$. Here,~$\hat{\boldsymbol{Q}}$ denotes a quadrature operator vector for a two-mode continuous variable quantum system and written as~$\hat{\boldsymbol{Q}}=(\hat{x}_1,\hat{p}_{1},\hat{x}_{2},\hat{p}_{2})^\text{T}$ satisfying the canonical commutation relation,~$[\hat{Q}_j,\hat{Q}_k]=i\Omega_{jk}$, where~$\Omega=\begin{pmatrix} 0 & 1 \\ -1 & 0 \end{pmatrix} \times \mathbb{I}_2$ and ~$\mathbb{I}_n$ is the~$n\times n$ identity matrix.

For the output state~$\hat{\rho}_{\boldsymbol{T}}$ for the TMSV state input, it can be shown that~$\boldsymbol{d}=(0,0,0,0)^\text{T}$, while 
\begin{align}
\boldsymbol{V}(\boldsymbol{T})= \begin{pmatrix}
v_1 & 0 & -v_3 & 0 \\
0 & v_1 & 0 & v_3\\
-v_3 & 0 & v_2 & 0 \\
0 & v_3 & 0  & v_2
\end{pmatrix}
\end{align}
where 
\begin{align}
v_1&=\frac{1}{2}+ \eta_\text{L} T_\text{L}  \sinh^2 r,\\
v_2&=\frac{1}{2}+ \eta_\text{R} T_\text{R} \sinh^2 r,\\
v_3&=\frac{1}{2}\sqrt{\eta_\text{L}\eta_\text{R}T_\text{L} T_\text{R}}\sinh 2r,
\end{align}
where a squeezing parameter has been assumed to be real, i.e.,~$\xi=r\in\mathbb{R}$.

For two states described by the covariance matrices~$\boldsymbol{V}_1$ and~$\boldsymbol{V}_2$ but having zero displacement, the quantum fidelity can be written as~\cite{Marian2012, Banchi2015}
\begin{align}
{\cal F}(\boldsymbol{V}_1,\boldsymbol{V}_2)=\left[\sqrt{\gamma}+\sqrt{\lambda}-\sqrt{(\sqrt{\gamma}+\sqrt{\lambda})^2-\delta}\right]^{-1},
\end{align}
where
\begin{align}
\delta &= \text{det}(\boldsymbol{V}_1+\boldsymbol{V}_2),\\
\gamma &=16~\text{det}(\Omega \boldsymbol{V}_1 \Omega \boldsymbol{V}_2 -\mathbb{I}_4/4),\\
\lambda &=16~\text{det}(\boldsymbol{V}_1+i\Omega/2)\text{det}(\boldsymbol{V}_2+i\Omega/2).
\end{align}
Upon with the above formalism and analytical form of the quantum fidelity, one can thus derive the QFIM of Eqs.~\eqref{eq:QFIM_TMSV_diagonal} and~\eqref{eq:QFIM_TMSV_off-diagonal} in the main text.

\bibliography{reference.bib}

\end{document}